\documentclass[10pt]{elsart}
\usepackage{natbib,epsfig,graphics,amssymb,amsfonts,amsmath,bbm}
\usepackage[latin1]{inputenc}
\usepackage{verbatim}
\newtheorem{theo}{Theorem}

\newtheorem{corol}[theo]{Corollary}

\newenvironment{proof}{\noindent\textit{Proof.}}{\hfill $\Box$\smallskip}

\def\cal{\mathcal}

\def\R{{\mathbb R}}
\def\E{{\mathbb E}}

\def\P{{\mathbb P}}
\def\ind{\mathbbm{1}}
\def\Var{\mathrm{Var}}

\def\eps{\varepsilon}
\def\etal{{\em et al.}}

\def\wT{B}
\def\pT{\widetilde{B}^\eps}
\def\wB{B}
\def\wL{L}
\def\pL{\widetilde{L}}
\def\wA{{\cal A}}

\def\wAc{\bar{{\cal A}}}

\def\pL{\widetilde{L}^\eps}
\def\pT{\widetilde{B}^\eps}
\def\pA{\widetilde{{\cal A}}^\eps}

\begin{document}

\begin{frontmatter}

\title{Integration of streaming services and TCP data transmission in the Internet}
\author[INRIA]{N. Antunes}
\author[INRIA]{C. Fricker}
\author[FT]{F. Guillemin}
\author[INRIA]{Ph. Robert}

\address[INRIA]{INRIA,  RAP project, Domaine de Voluceau, 78153 Le Chesnay, France}
\address[FT]{France Telecom, Division R\&D, 22300 Lannion, France}

\begin{abstract}
We study in this paper the integration of elastic and streaming traffic on a same link in
an IP network. We are specifically interested in the computation of the mean bit rate
obtained by a data transfer. For this purpose, we consider that the bit rate offered by
streaming traffic is low, of the order of magnitude of a small parameter $\eps \ll 1$ and
related to  an auxiliary stationary Markovian process $(X(t))$. Under the
assumption that data transfers are exponentially distributed, arrive according to a
Poisson process, and  share the available bandwidth according to the ideal processor
sharing discipline, we derive the mean bit rate of a data transfer as a power series
expansion in $\eps$. Since the system can be described by means of an $M/M/1$ queue with a
time-varying server rate, which depends upon the parameter $\eps$ and process
$(X(t))$, the key issue is to compute an expansion of the area swept under the occupation
process of this queue in a busy period.  We obtain closed formulas for the power series
expansion in $\eps$ of the mean bit rate, which allow us to verify the validity of the
so-called reduced service rate  at the first order. The second order term yields more
insight into the negative impact of the variability of streaming flows.  
\end{abstract}


\end{frontmatter}

\section{Introduction}
The emergence of the Internet as the universal multi-service network raises major traffic
engineering problems, in particular with regard  to the coexistence on the same
transmission links of real time and data services. As a matter of fact, these two types of
services have different requirements in terms of transfer delay and loss, data
transmission being very  sensitive to packet loss but  relatively tolerant to delay
whereas real time services have strict transfer delay constraints. While classical data
transfers are usually controlled by TCP (Transmission Control Protocol), which aims at
achieving a fair bandwidth allocation at a bottleneck link (see  Massoulié and
Roberts~\cite{Massoulie} for a discussion on modeling TCP at the flow level and processor
sharing), real time services most of the time are supported by the unreliable UDP
protocol, even if some transmission control can be performed by upper layers (e.g.,
RTCP). Real time services  thus reduce the transmission capacity for   data transfers. 

This problem has been addressed by Delcoigne \etal\ \cite{Proutiere}, where stochastic
bounds have been obtained for the bit rate seen by a TCP data transfer, when elastic
traffic and unresponsive streaming flows are  multiplexed on a same link (see also Bonald
and Proutière~\cite{Bonald:01}). From a theoretical point of view, this problem can be
seen as the analysis of a priority system, where streaming flows have priority over data
traffic. In this context, a usual approximation (referred to as Reduced Service Rate, RSR)
consists of assuming that everything happens as if the service rate for data were reduced
by the  mean bit rate offered by streaming flows. This approximation has been investigated
for the number of active data flows by Antunes \etal\  \cite{Nelson}, when the load
offered by streaming flows is very small. We note that the same kind of problem has been
addressed in the technical literature by N\'u\~{n}ez-Queija and Boxma \cite{Nunez1} in the
context of ABR service in ATM networks and more recently by N\'u\~{n}ez-Queija
\cite{Nunez2,Nunez3} via  matrix analysis for systems described by means of quasi birth
and death processes. In a similar context, N\'u\~nez-Queija \etal\ \cite{Altman} use a
perturbation technique for studying a priority system, where priority traffic offers a
small load. Although the systems considered in this paper are quite general, no explicit
expressions for the terms of the expansions are provided. 
Finally, note that systems with different speeds are also of interest for
analyzing the coexistence of different traffic types \cite{Kurkova}. 

In this paper, we  investigate the mean bit rate obtained by  a data transfer when elastic
traffic  and unresponsive streaming  flows are  multiplexed on  a same  transmission link.
Along the same line of investigations  as Antunes \etal~\cite{Nelson}, because of the {\em
real difficulty} of the problems, the mean  load offered by streaming flows is supposed to
be  very  small  (controlled  by  a  parameter  $\varepsilon \ll  1$)  and a
perturbation analysis  for the analysis  of the  mean bit rate is done.  It is assumed
that elastic flows
arrive according  to a Poisson process and  share the available  bandwidth according to
the processor  sharing discipline.  In addition, to  simplify the computations,  we assume
that  the  service  time required  by  data  transfers  is exponentially  distributed  with
parameter $\mu$. Thus,  we have to deal  with an $M/M/1$ queue with  a time-varying server
rate, which depends upon the instantaneous number of active streaming flows. The
exponential distribution of the service time is of course not realistic in practice, but
as it will be seen later, the mathematical analysis is already very difficult in the case
so that our work should be considered as a first step in this domain. 

To compute the mean bit rate of a data transfer, we consider the quantity $\wA =
\int_0^{B} L(s) ds$, where $B$ is the length of the busy period and $L(t)$ is the number of customers at
time $t$ in the $M/M/1$ queue under consideration. The quantity $\wA$ is equal to the
cumulative waiting time in the $M/M/1$ queue and also represents the amount of data served
during a busy period. In the case of the $M/M/1$ PS sharing queue, if $\E(d)$ represents
the mean bit rate obtained by a data transfer, we have $\E(\wA) =
\E(N)\E(S)/\E(d)=\E(B)/\E(d)$, where $\E(S)$ is the mean service time (equal to $1/\mu$)
and $\E(N)$ is the mean number of customers served in a busy period. Thus, the computation
of $\E(\wA)$ allows us to estimate $\E(d)$, since the quantity $\E(B)$ has been computed
by Antunes \etal\ \cite{Nelson} as a power series expansion of $\varepsilon$. Note that in
the case of a classical $M/M/1$ queue, we have $E(B) = 1/(\mu (1-\rho))$ and
$\E(\wA)=1/(\mu(1-\rho)^2)$, which yields $\E(d) = (1-\rho)$, where $\rho$ is the
offered load. This is the classical result for an $M/G/1$ PS queue, which states that the
mean bit rate obtained by a data transfer is $(1-\rho)$ times the server rate (taken as
unity in this paper); see Massouli\'e and Roberts \cite{Massoulie}.  

In this paper, we derive a power series expansion in $\varepsilon$ of the quantity
$\E(\wA)$ in the case of an $M/M/1$ queue, whose server rate is modulated by an
auxiliary process $(X(t))$. We specifically assume that the server rate at time $t$  is
$\mu+\varepsilon p(X(t))$ for some function $p$ satisfying regularity  assumptions, the
process $(X(t))$ being stationary, ergodic,  and Markovian. The objective of this paper is,
first to check the validity of the RSR approximation, which claims that everything happens as
if the server rate were frozen at the value $\mu+\varepsilon\E[p(X(0))]$ and, second,
to get some qualitative insight on the impact of the variability of streaming flows on
elastic traffic. 

The   organization  of   this  paper   is   as  follows:   The  model   is  described   in
Section~\ref{model}, where the main result concerns the power series expansion in $\eps$
of $\E(\wA)$, which is the key quantity  for computing the mean bit rate of a data transfer.
In Section~\ref{applications}, the main result is applied to obtain the expansion of the mean
bit rate  of a  data transfer and  to analyze different  special cases.  Some concluding
remarks are presented in Section~\ref{conclusion}. The  quite technical 
proof of the main result is sketched in the Appendix.

\section{Model description}
\label{model}
Throughout this paper we consider a stable  $M/M/1$ queue with arrival rate $\lambda$ and service rate $\mu$; the load $\rho = \lambda/\mu<1$.
Let  $\wL(t)$ denote  the number of
customers at time $t$. The invariant distribution $\pi$ of $(\wL(t))$ is geometrically
distributed with parameter $\rho$. 

Let $\wT$ denote the duration of a busy period starting with one customer, that is, $\wT=\inf\{s\geq 0: \wL(s)=0\}$, 
given $\wL(0)=1$.  For $x\geq 1$, let $\wB_x$ denote the duration of a busy
period starting with $x$ customers. Note that  $\wB_1\stackrel{\text{dist.}}{=}
\wT$. In the following, when the variables $B$, $B_1$ and $B_1'$ are used in the
same expression, they are assumed to be independent with the same distribution as
$B$.

The quantity $\mathcal{A}$ defined in the Introduction represents the area swept under the
occupation process in a busy period. When several busy cycles are  considered, the
notation $\mathcal{A}_B$ will be used to indicate that the area is calculated for the
corresponding busy period of length $B$.  By definition, the relation $\mathcal{A}\geq B$
holds, the excess will be denoted by
$\wAc\stackrel{\text{def.}}{=}\wAc_\wT\stackrel{\text{def.}}{=}\mathcal{A}_\wT-\wT$. 
This queue will be referred to as the standard queue denoted, for short, by S-Queue.   

Streaming flows impact data transfers by  reducing the amount of available bandwidth. This
situation is  described by introducing  an $M/M/1$ queue  with arrival rate  $\lambda$ and
varying service rate driven by an ergodic  Markov process $(X(t))$ taking values in a state
space ${\cal S}$. Typically, the state space of the environment is a finite, countable set
when $(X(t))$  is a Markov  Modulated Poisson Process  or ${\cal S}=\R$  in the case  of a
diffusion,     for    instance    an     Ornstein-Uhlenbeck    process     (see    Fricker
\etal~\cite{Fricker:10}).  The invariant  measure of  the process  $(X(t))$ is  denoted by
$\nu$. The Markovian notation $\E_{x}(\cdot)$ will  refer only to the initial state $x$ of
the Markov process $(X(t))$.

Let $\pL(t)$ be the number of customers of the queue at time $t$.  The process
$(\pL(t),X(t))$ is a Markov process. If $X(t)=x$ and $L(t)=n>0$, then \emph{the service rate is given by } $\mu + \eps p(x)$
for some function $p(x)$ on the state space of the
environment ${\cal S}$ and some small parameter $\eps \geq 0$.    
For $t\geq 0$, let us define the quantities $p^+(t)=\max(p(t),0)$ and $ p^-(t)=\max(-p(t),0)$ so that $p(t)=p^+(t)-p^-(t)$. At time $t$,  the
additional capacity is therefore $\eps p^+(X(t))$ and  $\eps p^-(X(t))$ is the capacity
lost.

In the rest of this paper, we make the two following assumptions:\\
\centerline{{\em the function $p(x)$ is bounded ($H_1$)} \hfill and \hfill $\lambda+\eps\sup(|p(x)|: x\in{\cal S})<\mu$  ($H_2$).}
If assumption $(H_2)$ holds (it is a suffifient but not necessary condition) then the queue is stable and,  in this
case, the duration $\pT$ of a busy period
starting with one customer, ${\pT}=\inf\{s\geq 0: {\pL}(s)=0~ |~\pL(0)=1\}$, is a.s. finite. 
The queue  with time-varying  service rate  as defined above  will be  referred to  as the
perturbated  queue,  denoted,  for  short,   by  P-Queue.   The  case  $\eps=0$  obviously
corresponds to the S-Queue. 
The area for the perturbated queue over a busy cycle is defined as
\begin{equation*}
\pA=\int_0^{\pT} \pL(s)\,ds.
\end{equation*}
The basic idea of the perturbation analysis carried out in this paper for the quantity
$\pA$ defined by the above equation is to construct a
coupling between  the busy periods of the processes $(\pL(t))$ and $(L(t))$.  Provided that for both queues the arrival process is the same Poisson process
with parameter $\lambda$, we add and  remove departures as follows.
\begin{description}
\item[Additional departures.] When $p^+(X(t))>0$,  there is additional capacity when compared with the S-Queue and more departures can take place. These additional departures are counted by means of a point Process ${\cal N}^+=(t_i^+)$, with  $0< t_1^+\leq  t_2^+\leq \cdots$,  which is a
non-homogeneous Poisson process on $\R_+$ with  intensity given by $t\to \eps
p^+(X(t))$. Conditionally on $(X(t))$, the number of points of ${\cal N}^+$ in the
interval $[a,b]$  is Poisson with parameter $\eps\int_a^b  p^+(X(s))\,ds$. 
In particular the distribution of the location $ t_1^+ \geq 0$ of the first point of ${\cal N}^+$
after $0$ is given, for $x\geq 0$, by
\begin{equation}\label{eqt1}
\P( t_1^+\geq x)= \P({\cal N}^+([0,x])=0)=\E\left[\exp\left(-\eps \int_0^x  p^+(X(s))
\,ds\right)\right].
\end{equation}
\item[Removing Departures] On the other hand, when $p^-(X(t))>0$, the server rate is smaller  than in the S-queue. Let ${\cal  N}_\mu=(t_i)$, a Poisson process
with intensity  $\mu$ on $\R_+$ which  represents the non-decreasing  sequence of instants
when the customer of the S-queue (if not empty) may leave the queue.  We denote by ${\cal
N}^-$ the  point process obtained  as follows: For  $s>0$, a point  at $s$ of  the Poisson
process  ${\cal N}_\mu$ is  a point  of ${\cal  N}^-$ with  probability $\eps  p^-(X(s)) /
\mu$. (Note that this number is $\leq 1$ by assumption~($H_2$).)  
The point process ${\cal N}^-$  is Poisson  with intensity  $s\to \eps
p^-(X(s)) $. A point  of ${\cal N}^-$ is called a marked  departure.  The points of ${\cal
N}^-$ are denoted by $0< t_1^-\leq t_2^-\leq \cdots \leq t_n^- \leq \cdots$. By
definition,
\begin{equation}\label{eqtb1}
\P( t_1^-\geq x)=\E\left(\prod_{t_i, t_i\leq x}\left(1-\frac{\eps p^-(
  X(t_i))}{\mu}\right)\right), \quad x\geq 0.
\end{equation}
\end{description}

The processes defined above are non homogeneous Poisson processes (see Grandell~\cite{Grandell:01} for an account on this topic).
The main result of this paper is the  expansion in power series of $\varepsilon$ up to the second order of  $\E(\pA)$.

\begin{theo}
\label{main-result}
The second order expansion of the area swept under the occupation process of  the perturbated
queue during a busy period is given by 
\begin{equation}
\E\left(\pA\right) = \E\left(\wA\right) -\eps\frac{\E[p(X(0))](\lambda+\mu)}{(\mu-\lambda)^3}
- \eps^2(a_+ + a_- + a_\pm) +o(\eps^2),
\end{equation}
where the coefficients $a_+$, $a_-$, $a_\pm$ are defined below by Equations~\eqref{exp-2} , \eqref{exp+2}, and~\eqref{exp+-}, respectively. 
\end{theo}

\section{Applications}
\label{applications}
In this section, as an application of Theorem~\ref{main-result}, we evaluate the mean bit
rate $\E(d_\eps)$ obtained  by an elastic data transfer when the server rate is
perturbated by the presence of streaming flows (coming through the term $\eps p(X(t))$ in
the service rate of the perturbated $M/M/1$ queue, equal to $\mu+\eps p(X(t))$). As
mentioned in the Introduction, we have $\E(\pA) = \E(\widetilde{B}^\eps)/\E(d_\eps)$. The
average of the duration $\widetilde{B}^\eps$ of the corresponding busy period has been
studied by Antunes \etal\  \cite{Nelson} and can be expanded in power series of $\eps$ as
follows.

\begin{theo}
The expansion of $\E (\pT)$, the mean duration of a busy period, is given by
$\E (\pT) =  {1}/{(\mu-\lambda)} -\eps {\E_\nu(p(X(0)))}/{(\mu-\lambda)^2}  + (b_-
- b_+) \eps^2 + o(\eps^2). $
where $b_+$ and $b_-$ are given by,  with the notation of Proposition~\ref{prop2},
\begin{align}
b_+ &=  \ - \frac{1}{\mu}\E\left(\int_{0}^{\wT}(\wT-v)\,
\E_\nu\left(p^+(X(0))p^+(X(v))\right)\,dv\right)  \notag \\
&\qquad\qquad  -\frac{1}{\mu^2(1-\rho)}  \E\left(\sum_{i=1}^{H} \sum_{j=1}^{N_i}\int_{0}^{A_i} p^+(X(u)) p^-(X(D_i^j))\,du \right), \label{defb+}\\
b_- &= \frac {1 }{\mu^2(1-\rho)} \left ( -  \E \left ( \sum_{i=1}^N   \int_0^{\wT + \wB_1}
p^- (X(D_i))  p^+ (X(s))  \, ds \right  ) \right. \notag \\ 
&\qquad\qquad\left.+\frac{1}{\mu} \E\left (\sum_{i=1}^{N} \sum_{k=1}^{N^{\prime}}   p^- (X(D_i)) p^- (X(\wT + D_k^{\prime} ))\right ) \right ).\label{defb-}
\end{align}
\end{theo}
By using Equations~\eqref{defb+} and~\eqref{defb-} and Theorem~\ref{main-result},
straightforward computations show that the quantity $\E(d_\eps)$ can then expanded in power 
series of $\eps$ as follows. 
\begin{prop}
The mean bit rate of an elastic data transfer can be expanded in power series of $\eps$ as 
\begin{equation}
\label{meanDeps}
\E(d_\eps) = 1-\rho +\frac{\rho\E_\nu(p(X(0)))}{\mu}\eps + c \eps^2+o(\eps^2),
\end{equation}
where the coefficient $c$ is given by
\begin{equation}
\label{defc}
c = \E_\nu(p(X(0)))^2\frac{\rho(1+\rho)}{\mu^2(1-\rho)} + \mu(1-\rho)^2\left((1-\rho)(a_++a_-+a_\pm)+b_-- b_+\right),
\end{equation}
the quantities $a_+$, $a_-$, $a_\pm$, $b_+$ and $b_-$ being defined by equations~\eqref{exp-2}, \eqref{exp+2}, \eqref{exp+-},  \eqref{defb+} and \eqref{defb-}, respectively.
\end{prop}
From Equation~\eqref{meanDeps}, we immediately deduce that as far as the first order term
is concerned, the RSR approximation is valid:
for  an   $M/M/1$  queue   with   service  rate
$\mu+\eps\E_\nu(p(X(0)))$,  the mean bit  rate  denoted by $\hat{d}$ obtained  by a
customer is   given by 
$
\E(\hat{d}) =
1-{\lambda}/{\left(\mu+\eps\E_\nu[p(X(0))]\right)} =  1-\rho +\eps{\rho}\E_\nu[p(X(0))]/{\mu}   +o(\eps)$
Unfortunately, the  coefficient $c$  defined by  Equation~\eqref{defc} intricately
depends upon the correlation structure of the modulating process $(X(t))$ and the dynamics
of the $M/M/1$ queue. Because of this complexity, three cases  of practical interest in the
following are considered: non-positive  perturbation functions, non-negative  perturbation
functions, and special environments (namely, fast and slow environments).

\vspace{-9mm}

\subsection{Non-positive Perturbation Functions}\vspace{-9mm}
We assume in this section that the perturbation function is  non-positive so that the
environment  uses a part of the capacity  of  the $M/M/1$  queue with  constant service  rate
$\mu$. This application is motivated by  the following practical situation. Coming back to
the coexistence of elastic and streaming  traffic in the Internet, assume that priority is
given  to streaming  traffic in  a buffer  of a  router. The  bandwidth available  for
non-priority  traffic  is  the  transmission  link  reduced  by  the  rate of
streaming  traffic. Denoting  by $\eps r(X(t))$  the rate of streaming  traffic at time  $t$ (for
instance $\eps$ may represent the peak rate of a streaming flow and $r(X(t))$ the number of
such flows active  at time $t$), the  service rate available for non-priority traffic is
$\mu -  \eps r(X(t))$.    Setting $p(x) = - r(x)$, the function $p(x)$ is non-positive.  
\begin{prop}\label{cheval}
When $p^+ \equiv 0$, if $\hat{d}$ is the mean bit rate of the $M/M/1$ queue with service
rate $\mu+\eps\E[p(X(0))]$, then  with the notation of Proposition~\ref{prop2},
\begin{multline}
\label{dif1}
\lim_{\eps\to 0}\frac{1}{\eps^2}\E\left(d_\eps -\hat{d} \right)=  -\frac{(1-\rho)^2}{\mu^2}\E\left(\sum_{1\leq i<j\leq N}  C_p\left( X(D_j  -D_i)\right)\right)\\
\\ -\frac{(1-\rho)^3}{\mu}\E\left(\sum_{i=1}^{N}\sum_{j=1}^{N'}  C_p( \wT  -D_i + D_j^{\prime}) \left (B_1-D_j' \right ) \right),
\end{multline}
where the function $C_p(u)$ is defined for $u\geq 0$ by
\begin{equation}
\label{defCpu}
C_p(u)=\E_\nu\left[p(X(0))p(X(u))\right]-\E_\nu\left[p(X(0))\right]^2 
\end{equation}
and is, up to the factor $\eps^2$, the auto-covariance function of the variable capacity of
the perturbated queue.  
\end{prop}
The above result shows that if the process $(X(t))$ is positively correlated,
i.e. $C_p(\cdot)\geq 0$, then $\lim_{\eps\to 0}\E\left(d_\eps -\hat{d}
\right)/\eps^2<0$. The environment has therefore a negative impact on the performances of the
system in this case. 

\begin{proof}
The terms $a_+$, $a_\pm$ and $b_+$ in the coefficient $c$ defined by Equation~\eqref{defc}
are equal to 0. In addition,  one easily checks that
\begin{multline*}
(1-\rho)a_- +b_- = -\frac {1 }{\mu^3}  \E \left ( \sum_{1\leq i<j\leq  N}   p^- (X(D_i))  p^- (X(D_j))\right  ) \\ 
-\frac{(1-\rho)}{\mu^2} \E\left (\sum_{i=1}^{N} \sum_{k=1}^{N^{\prime}}   p^- (X(D_i)) p^- (X(\wT + D_k^{\prime} )) (B_1-D_j') \right ).
\end{multline*}
One concludes by using Expansion~\eqref{meanDeps}.
\end{proof}
\vspace{-9mm}

\subsection{Non-negative Perturbation Functions}

\vspace{-9mm}
It is assumed in this section that $p^-\equiv 0$. We have the following result, which  is the
analogue of Proposition~\ref{cheval} for this case.  Contrary to the above proposition,
the expansion has a more explicit expression. Its (straightforward)  proof  is omitted.
\begin{prop}\label{cheval2}
When $p^-\equiv 0$, with the same notations as in  Proposition~\ref{cheval}.
\begin{multline}
\lim_{\eps\to 0}\frac{1}{\eps^2}\E\left(d_\eps -\hat{d} \right)  
=(1-\rho)^3 \left(   \E\left(\int_{0}^{\wT} (\wT-v)\,C_p(v)\,dv\right)\right. \\
\left. -\frac{(1-\rho)\mu}{2}\E\left(\int_{0}^{\wT} (\wT-v)^2\,C_p(v)\,dv\right) \right).
\end{multline}
\end{prop}
An integration by parts and some calculations give the following corollary.
\begin{corol}
When the correlation function of the
environment is exponentially decreasing, i.e., when $C_p(x)=\Var[p(X(0))]\,e^{-\alpha x}$ for all $x\geq 0$ and  some $\alpha >0$, then
\begin{equation}\label{auxey}
\lim_{\eps\to 0} \frac{1}{\eps^2}\,\E\left(d_\eps -\hat{d} \right) = 
\frac{1}{\mu^2}\left(\E\left(e^{-\alpha B^{**}}\right) - 
\frac{1+\rho}{1-\rho}\E\left(e^{-\alpha B^{***}}\right)\right),
\end{equation}
with the convention that, if $Z$ is some integrable non-negative random variable, the
density on $\R_+$ of the variable $Z^*$ is defined as 
$$
\P\left(Z^*\geq x\right)= \P(Z\geq x)/{\E(Z)},$$ for $x\geq 0$,
and $Z^{**}$ stands for $(Z^*)^*$, and $Z^{***}$ for $(Z^{**})^*$.
\end{corol}
When $\alpha$ is small, the right hand side of Equation~\eqref{auxey} is equivalent to the
quantity $-2\rho/\mu^2 <0$. This shows that non-negative perturbation functions have a
negative impact at the second order on the mean bit rate of elastic data transfers.
 
\vspace{-5mm}
{\bf Remark.} This result could maybe be extended to a perturbated function with a non-constant sign but it
is out of reach because of the complexity of the secund order term.

\subsection{Fast and slow Environments}

The performance  of the system  in two limit  regimes, called fast  and slow
environments, are now evaluated. These  regimes are very  useful, since performance  in the
limit  regimes is insensitive and  only depends  on appropriately defined
parameters. Such a  situation has also been analyzed by Delcoigne \emph{et
  al.}~\cite{Proutiere} through stochastic bounds.  

The environment is scaled by a factor $\alpha>0$, such that at time $t$ the environment is
supposed to be $X(\alpha t)$. The behavior when $\alpha$ goes to infinity and zero is
investigated.  

When the parameter $\alpha$ is very large, the environment process approximately averages
the capacity of the variable queue. For a large $\alpha$ and for $t$ and $h>0$, the total service capacity available during $t$ and $t+h$ is given by
\[
\mu h +\eps \int_{t}^{t+h} p(X(\alpha u))\,du \stackrel{\text{dist.}}{=}
\mu h +\eps \frac{1}{\alpha}\int_{0}^{\alpha h} p(X(u))\,du\sim (\mu+\eps\E[p(X(0))])h
\]
using the stationarity of $(X(t))$ and the ergodic theorem. Thus, when $\alpha$ tends to infinity, the variations completely vanish and the service rate reduces to a constant.

On the other hand, for small values of $\alpha$, the environment process remains almost constant over  the busy period of the P-Queue. As  $\alpha$ goes to $0$, the variation disappears and the environment is frozen in the initial state of the process: the service rate is constant and equal to $\mu+\eps p(X(0))$.

This intuitive picture is rigorously established in the next proposition. 
In the following a general perturbation function $p$ is considered together with some stationary Markov process $(X(t))$ with invariant probability distribution $\nu$. It is assumed that it
verifies a mixing condition such as
\begin{equation}\label{mixing}
\lim_{t\to+\infty} |\E[f(X(0)g(X(t))]-\E_\nu(f)\E_\nu(g)|=0
\end{equation}
for any Borelian bounded functions $f$ and $g$ on the state space ${\cal S}$. Note that
this condition is not restrictive in general since it is true for any ergodic Markov process
with a countable (or finite) state space or for any diffusion on $\R^d$.

Under the above assumptions, we have the following result; the proof relies on the use of the mixing condition~\eqref{mixing} and can be found in the paper by Antunes \etal\ \cite{Nelson2}.

\begin{prop}\label{A_limit}
When the environment is given by $(X(\alpha t))$ and Relation~\eqref{mixing} holds, then when $\eps$ tends to 0,
$
\E(d_\eps)-\E(\hat{d}) =  \Psi(\alpha)\eps^2 +o(\eps^2)
$
where
$$
\lim_{\alpha\to +\infty}  \Psi(\alpha) = -\rho \frac{\E_\nu(p(X(0)))^2 }{\mu^2}
 \; \mbox{and} \;
\lim_{\alpha\to 0}  \Psi(\alpha) =-\rho \frac{\E_\nu\left(p(X(0))^2\right)}{\mu^2}.
$$
\end{prop}

The fast and slow  environment provide an explicit estimate of the second order term,
where the slow  environment yields a worst  performance of the perturbated queue. It is
not clear that these limit regimes give a lower and upper bound of the performance of the
queue.

\section{Conclusion}
\label{conclusion}
We have investigated in this paper the impact on the performance of elastic data transfers
of the presence of streaming flows, when both kinds of traffic are multiplexed on a same
link of an IP network. By assuming that the perturbation due to streaming flows is of
small magnitude, a perturbation analysis can be performed in order to obtain
\emph{explicit} results for the mean bit rate achieved by a data transfer, under the
assumption that elastic streams share the available bandwidth according to the processor
sharing discipline. It turns out that at the first order, the so-called RSR approximation
is valid. This is not the case for the second order, for which the variability of
streaming flows seem to have a negative impact, at least for the three cases examined
here. 

Further investigations are needed in order to estimate the degradation suffered by data
transfers at the second order. The perturbation analysis carried out in this paper is
possible, because we have assumed that streaming flows offer a very small contribution to
the total load. When this is not the case, new tools have to be developed to estimate the
quality of data transfers.

\appendix
\section{Appendix: Proof of Theorem~\ref{main-result}}
\label{proof}

Since the derivation of the expansion of the average area  is somewhat technical, we begin
with the simplest case, which is the first order expansion.  In the following, we set
$\Delta^\eps=\E(\wA)-\E(\pA)$.  

\subsection{First order term}

We first consider an additional departure. Define ${\cal E}_{+} = \{ t_1^+ \leq \wT,
t_1^- \geq t_1^+ + \wB_{L( t_1^+)-1} \}$, where $\wB_{L( t_1^+)-1}$ is the duration
between the $t_1^+$ and the first time when the S-Queue has one customer (see Figure~\ref{fig1}-(b)). On this event, an additional departure is added and the busy period of the P-Queue
finishes before a departure is canceled. For the first order term of the expansion of the
mean value of $\Delta^\eps$ on the event ${\cal E}_{+}$, we only need to consider the case
where there is only one  additional departure during the busy period of the P-Queue. The
probability that two additional jumps occur in the same busy period is of the order of
magnitude of $\eps^2$ since the intensity of the associated Poisson process is
proportional to $\eps$. 

\begin{lem}
\label{lemme1}
In the case of a single additional departure
\begin{equation*}
\E \left(\Delta^\eps\ind_{{\cal E}_{+} } \right) =  \eps\frac{\E_\nu[p^+(X(0))](\lambda+\mu)}{(\mu-\lambda)^3}+o(\eps).
\end{equation*}
\end{lem}
\vspace{-5mm}
\begin{proof}
The difference $\Delta^\eps$ on the event $\{t_1^+ \leq \wT, \, t_2^+ > t_1^+ +
\wB_{L(t_1^+)-1}, \, t_1^- \geq t_1^+ + \wB_{L(t_1^+)-1}\}$ is  the sum of
two disjoint areas (see Figure~\ref{fig1}-(a)). The first  one is given by the distance between  $t_1^+$ and  the end of the busy period of the
S-Queue. By the strong Markov property at the stopping time $\pT$, conditionally on the
event $\{t_1^+ \leq \wT, t_2^+ \geq t_1^+  + \wB_{L(t_1^+)-1} , t_1^- \geq t_1^+ +
\wB_{L(t_1^+)-1} \}$, the S-Queue starts at time $\pT$ an independent busy period with one
customer (with duration $\wT_1$). The second area of $\Delta^\eps$ is then given by the
area of the {\em sub-busy periods in $\wT_1$}, i.e.  periods when $t{\to}L(t)$ is $>1$ in the second b.p.  

\begin{figure}[ht]
\scalebox{0.3}{\includegraphics{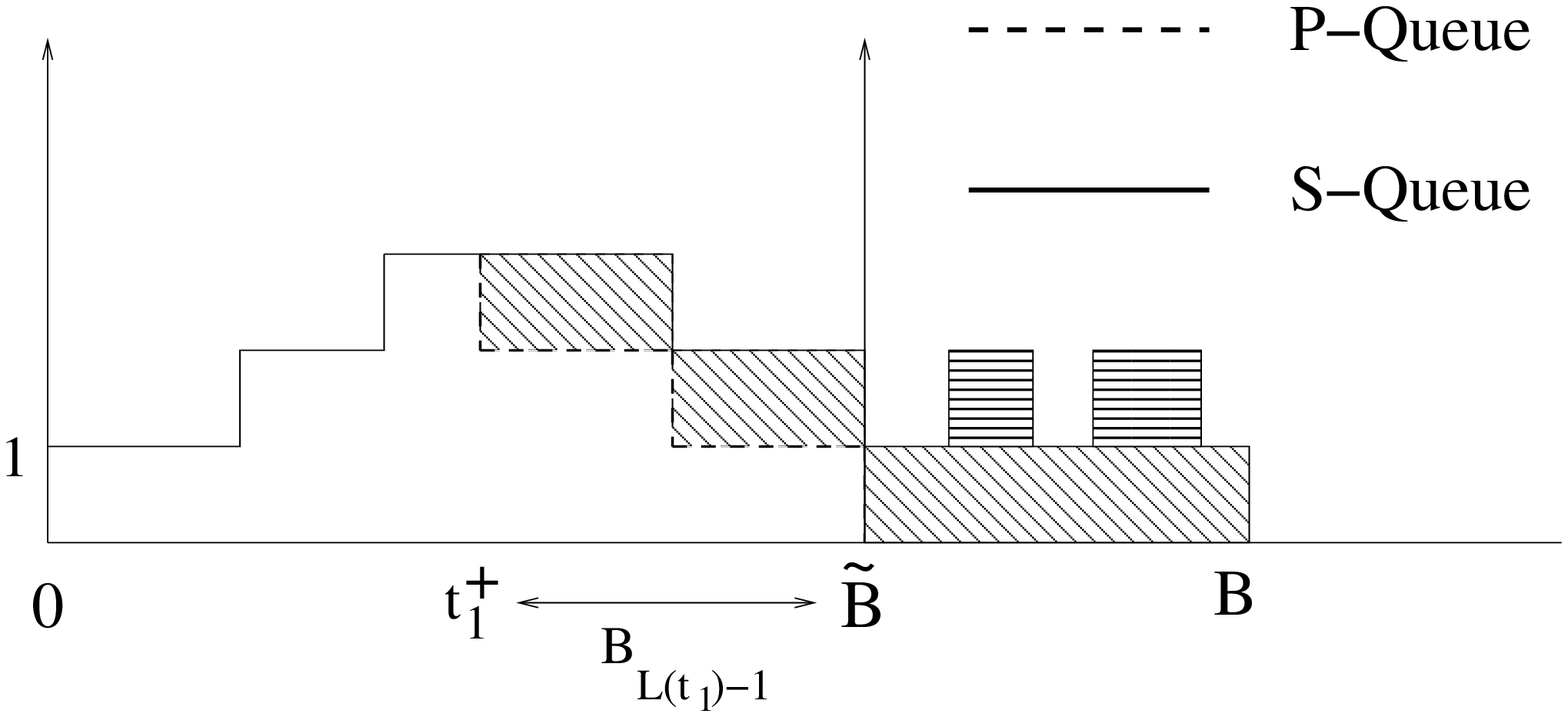}}
\hspace{5mm}\scalebox{0.3}{\includegraphics{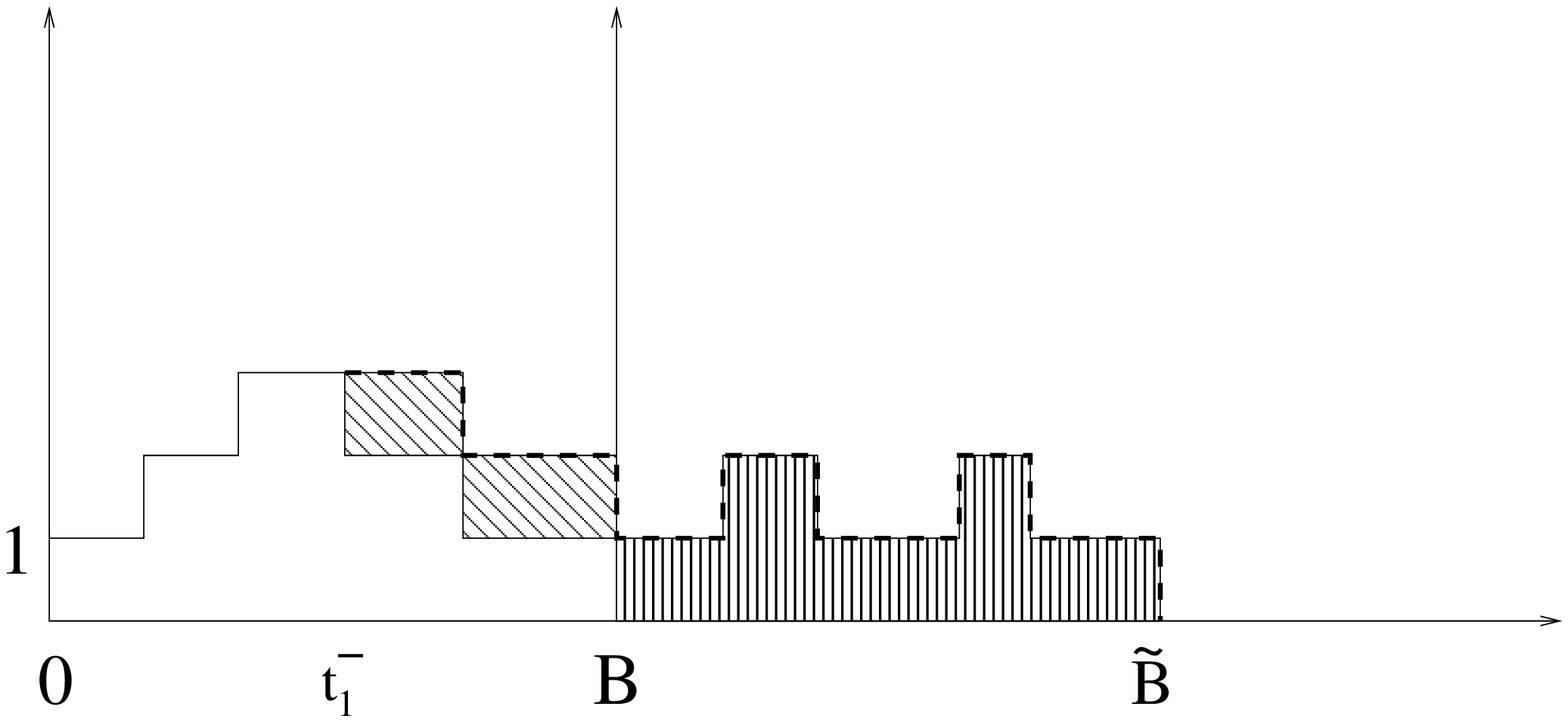}}
\label{fig1}
\caption{(a) One additional departure --- \vspace{10mm} (b) One marked departure}
\end{figure}
It follows that
\begin{multline*}\label{eps-1-add}
\E \left(\Delta^\eps\ind_{{\cal E}_{+} } \right) =\E \left( \Delta^\eps \ind_{\{t_1^+ \leq \wT, t_2^+ \geq t_1^+ + \wB_{L(t_1^+)-1} , t_1^- \geq
 t_1^+ + \wB_{L(t_1^+)-1} \}} \right ) +o(\eps)= \\
\E \left ((\wT- t_1^+) \ind_{\{t_1^+ \leq \wT, t_2^+ \geq t_1^+ + \wB_{L(t_1^+)-1} , t_1^- \geq
 t_1^+ + \wB_{L(t_1^+)-1} \}} \right )\\
+ \E(\wA_{B_1}-B_1) \P ( t_1^+ \leq \wT, t_2^+ \geq t_1^+  + \wB_{L(t_1^+)-1} , t_1^- \geq t_1^+ + \wB_{L(t_1^+)-1} )
\\= \P( t_1^+\leq \wT) \E(\wA_{B_1}-B_1) + \E \left((\wT-t_1^+) \ind_{\{ t_1^+\leq \wT\}} \right) + o(\eps).
\end{multline*}
Equation~\eqref{eqt1} and the boundedness of $p$ give that
\begin{multline*}
\P( t_1^+\leq \wT)=  1-\E\left[\exp\left(-\eps\int_0^{\wB} p^+(X(s))\,ds\right)\right] \\
= \eps \E\left[\int_0^{\wT} p^+(X(s))\,ds\right]+o(\eps) 
= \eps \E(\wT)\,\E\left[p^+(X(0))\right]+o(\eps) \\ = \frac{\eps}{\mu-\lambda}\E\left[p^+(X(0))\right]+o(\eps)
\end{multline*}
by independence between  $\wT$ and $(X(t))$ and by the stationarity of $(X(t))$. 

Similarly, 
\begin{align*}
\E \left((\wT-t_1^+) \ind_{\{ t_1^+\leq \wT\}} \right) &= \E\left(\int_0^{\wT} \eps p^+(X(u))e^{-\eps\int_0^{u}p^+(X(s))\,ds} (\wT-u) \,du \right )\\
&= \eps \E[p^+(X(0))]\frac{\E(\wT^2)}{2} + o(\eps)
\end{align*}
 Since $\E\left(\wT^2\right)=2/\mu^2(1-\rho)^3$ and
$\E(\wA_{\wB_1})=\mu/(\mu-\lambda)^2$ (see for instance standard books such as Cohen
 \cite{Cohen:01})). The lemma is proved.
\end{proof}

We now turn to the case, when there is one removed departure. On the event ${\cal E}_{-}=\{ t_1^-\leq \wT,\, \wT + \wB_1 \leq  t_1^+\}$, a marked departure occurs and no departures are added before the completion of the busy
period $\wT_1$. We derive the first order expansion of the mean value of $\Delta^\eps$ on ${\cal E}_{-}$.

Assume that there is only one marked departure and no additional jumps during the busy period of the P-Queue. In this case, at the end of the busy period of the S-Queue, the P-Queue has one customer and the difference  is the distance between 
$t_1^-$ and  the end of the busy period $\wT$.
At time $\wT$, the P-Queue starts a busy period with one customer and provided that there
are no marked and additional departures during $(\wT,\wT+\pT)$, the difference  has the
same distribution as the area of a busy period $\wB_1$ of the standard queue  (see Figure~\ref{fig1}-(b)).  

\begin{lem}
\label{lemme2}
In the case of a single marked  departure
\begin{equation*}
\E \left(\Delta^\eps\ind_{\cal{E}_-} \right) =  - \eps \frac{\E[p^-(X(0))](\lambda+\mu)}{(\mu-\lambda)^3}+o(\eps).
\end{equation*}
\end{lem}

\begin{proof}
By using the same arguments as before, one obtains the relation
\begin{multline} \label{mark1}
\E \left(\Delta^\eps\ind_{\cal{E}_-} \right) = \E \left (\Delta^\eps \ind_{\{ t_1^-\leq \wT,  t_2^- \geq   \wT +\wB_1  ,  t_1^+ \geq \wT +\wB_1  \}} \right )+o(\eps) \\ =  - \E\left( (\wT- t_1^- + \wA_{\wB_1}) \ind_{\{ t_1^-\leq \wT,  t_2^- \geq    \wT +\wB_1  ,  t_1^+ \geq \wT +\wB_1  \}}\right) +o(\eps)
\end{multline}
Hence, $ \E(\Delta^\eps  \ind_{{\cal E}_{-}} )= - \E\left( (\wT- t_1^- + \wA_{\wB_1})
\ind_{\{ t_1^-\leq \wT,  t_2^- \geq    \wT +\wB_1  ,  t_1^+ \geq \wT +\wB_1  \}}\right) +
o(\eps)$, so that 
$ \E(\Delta^\eps  \ind_{{\cal E}_{-}} )=- \E((\wT- t_1^-)\ind_{\{ t_1^- \leq \wT\}}) - \E(\wA_{\wB_1}) \P( t_1^- \leq \wT) + o(\eps). $

To estimate $\P(t_1^- \leq \wT )$, let $(D_i)$ denote the sequence of departures times and $N$ the number of customers served during the busy period of length $\wT$, then Equation~\eqref{eqtb1} gives the identity
\begin{align*}
\P(t_1^- \leq  \wT ) &= \E \left(  \sum_{i=1}^{N} \frac{\eps
  p^{-}(X(D_{i}))}{\mu} \prod_{j=1}^{i-1} \left ( 1- \frac{\eps p^-(X(D_j))}{\mu
} \right )
\right)\\
&=\frac{\eps}{\mu}  \E\left( \sum_{i=1}^{N}   p^-(X(D_{i})) \right) +o(\eps)
=\frac{\eps}{\mu}  \E(N)\E[p^-(X(D_{1}))] +o(\eps)
\end{align*}
by stationarity of $(X(t))$ and Wald's Formula with $\E(N)=1/(1-\rho)$.
Similarly, 
\begin{align*}
\E\left((\wT- t_1^-)\ind_{\{ t_1^-\leq \wT\}}\right) &=\frac{\eps}{\mu}  \E\left( \sum_{i=1}^{N}   p^-(X(D_{i})) (\wT-D_i) \right) + o(\eps)\\
&=\eps \frac{\E[p^-(X(D_{1}))]}{\mu} (\E(N\wT) - E(D)) 
\end{align*}
where $D=\sum_{i=1}^N D_i$ is the sum of the departures in the busy period of the S-Queue.
Using the fact that  $\E(D)=\mu^2/(\mu-\lambda)^3$, $\E(N\wT)=(1+\rho)/(\mu(1-\rho)^3)$
and $\E(\wA)=\mu/(\mu-\lambda)^2$, the  result is proved.
\end{proof}

Combining Lemmas~\ref{lemme1} and \ref{lemme2} yields the first order term indicated in Theorem~\ref{main-result}.
\subsection{Second order term}
To compute the second order term in the power series expansion in $\eps$ of
$\E\left(\Delta^\eps\right)$, three cases have to be considered:
\vspace{-5mm}
\begin{itemize}
\item[---] $t_1^+{\leq} \pT$ or $t_2^+{\leq} \pT$: one or two additional
departures occur in a busy period;
\item[---] $t_1^-\leq \pT$ or $t_2^-\leq \pT$: one or two departures are canceled;
\item[---] $t_1^+\leq \pT$ and $t_1^-\leq \pT$: one additional departure takes place and another one is canceled. 
\end{itemize}
\vspace{-5mm}
It is not difficult to show that any event involving a third jump yields a term of the
order $\eps^3$ in the expansion of the mean bit rate. 
Due to the space constraints, a  part of the expansion is proved. The complete
proofs of the expansion can be found in Antunes {\em et al.}~\cite{Nelson2}.

In the case that two additional departures occur during $\pT$, the difference between the
areas of the busy periods  due to the first additional jump is given by the two first
terms on the right hand side of the following equation
\begin{multline}\label{eps2}
\E \left ( \Delta^\eps \ind_{\{ t_1^+\leq \wT,  t_2^+ \leq  t_1^+ + \wB_{L( t_1^+)-1} \}} \right)=
\E \left( (\wT- t_1^+)  \ind_{\{ t_1^+\leq \wT,  t_2^+ \leq  t_1^+ + \wB_{L( t_1^+)-1} \}} \right ) \\
+ \E(\wAc_{\wB_1}) \P(t_1^+\leq \wT,  t_2^+ \leq  t_1^+ + \wB_{L( t_1^+)-1})
+ \E\left(\wB_{\wL( t_2^+)-1}\ind_{\{ t_1^+\leq \wT,  t_2^+ <  t_1^+ + \wB_{L( t_1^+)-1}\}} \right)\\
+ \E(\wAc_{B_1'}) \P(t_1^+\leq \wT,  t_2^+ <  t_1^+ + \wB_{L( t_1^+)-1}) + o(\eps^2),
\end{multline} 
which follows by the same arguments
stated for only one additional departure. 

\begin{figure}[ht]
\begin{center}\scalebox{0.5}{\includegraphics{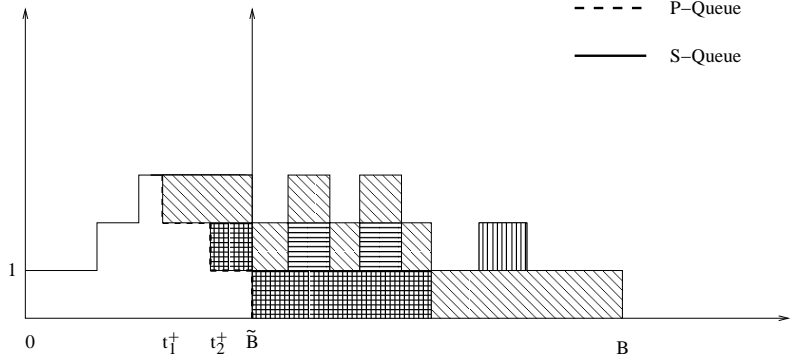}}\end{center}
\caption{Two additional departures}
\end{figure}

Due to the second additional jump, the difference $\Delta^\eps$ increases by the sum of 
two disjoint areas. The first one is given by  $\wB_{\wL( t_2^+)-1}$ and represents the
distance from the second additional jump until the  first time  the S-Queue with less than
one customer is empty. 
 Note that by conditioning on the event  $\{ t_1^+\leq \wT,  t_2^+ \leq  t_1^+ + \wB_{L(
   t_1^+)-1} \}$,  at the stopping time $\pT$, a new busy  period $B_1'$  
starts with the same distribution as $B$. Thus, the second area is given by the area of
sub-busy periods in $|wB_1$ (periods where $t\to L(t)$ is $>1$) in $\wB_1'$, hence
\begin{align}
\E\left(\Delta^\eps\ind_{{\cal E}_+} \right) &= 
\E \left( \Delta^\eps \ind_{\{t_1^+ \leq \wT, t_2^+ \geq t_1^+ + \wB_{L(t_1^+)-1} , t_1^-
  \geq t_1^+ + \wB_{L(t_1^+)-1} \}} \right )\notag \\
&+E \left ( \Delta^\eps \ind_{\{ t_1^+\leq \wT,  t_2^+ \leq  t_1^+ + \wB_{L( t_1^+)-1} \}}
\right) +o(\eps^2)\notag \\
 {=} \E\left( (B-t_1^+) \ind_{\{ t_1^+\leq \wT\}} \right)& {+} \E\left(\wB_{\wL(
   t_2^+)-1}\ind_{\{ t_1^+\leq \wT,  t_2^+ <  t_1^+ + \wB_{L( t_1^+)-1}\}} \right){+}
 \E(\wAc_{B_1})\P(  t_1^+ {<} \wT)\notag  \\  +\E(\wAc_{B_1'}) &\P(t_1^+\leq \wT,  t_2^+ <
 t_1^+ + \wB_{L( t_1^+)-1})+K(\eps) +o(\eps^2).   \label{eps2-2}
\end{align}
where $K(\eps)$ is a term which is not expressed here for sake of simplicity.
The following proposition gives the expansion of some of the terms of
Equation~\eqref{eps2-2}. The other expansions are done much in the same way (with various complications). 
\begin{prop}\label{prop+}
The following expansions hold
\begin{multline*}
\P( t_1^+{\leq} \wT){=} \\
\eps \frac{\E[p^+(X(0))]}{\mu-\lambda}{-}\eps^2\E\left(\int_{0}^{\wB}(\wB{-}v)\,
 \E\left[p^+(X(0))p^+(X(v))\right]\,dv\right){+}o(\eps^2),
\end{multline*}
\begin{multline*}
\P( t_1^+ {<} \wT,  t_2^+ {\leq}  t_1^+ {+}\wB_{\wL( t_1^+){-}1}) \\ {=} 
\eps^2 \rho \E\left(\int_{0}^{\wB}(\wB{-}v)\,
 \E\left[p^+(X(0))p^+(X(v))\right]\,dv\right)  {+}o(\eps^2).
\end{multline*}
\end{prop}

\begin{proof}
Since $\P( t_1^+\leq \wT)=\E\left(1-\exp\left(-\eps\int_0^{\wT} p^+(X(s))\,ds\right)\right)$, the
expansion in power series of $\eps$ has the first term ${\E(p^+(X(0)))}/(\mu-\lambda)$ and second term 
\begin{align*}
\frac{1}{2}\E\left(\left(\int_0^{\wT} p^+(X(s))ds\right)^2\right)&= \E\left(\int_{0\leq u\leq
  v\leq \wT} \E\left(p^+(X(0))p^+(X(v-u))\right)dudv\right)\\
&= \E\left(\int_{0}^{\wB}(\wB{-}v)\,
 \E\left[p^+(X(0))p^+(X(v))\right]\,dv\right)
\end{align*}
by stationarity of the process $(X(t))$. The first expansion is proved.

The event $\{t^+_1 \leq \wT\, , t^+_2 <t^+_1 + \wB_{L(t^+_1) -1}\}$ occurs only when
$t^+_1$ and $t^+_2$ are in  a
sub-busy period $[s_{i-1}+E_{i},s_i]$, for some $i\in\{1,\ldots ,H\}$, (a period where $L$ is always $>1$). The variables $E_i$ are
i.i.d. exponential with parameter $\lambda$, $\wB_1^i=s_i-s_{i-1}- E_{i-1}$ has the same 
distribution as $\wT$ and $H$ is geometrically distributed with parameter $\lambda/(\lambda
+\mu)$. The probability that the first two additional jumps  are in the $i$-th  sub-busy
period, is   
\begin{multline*}
 \E \left( \int_{s_{i-1}+E_i}^{s_i}  \eps p^+(X(u))
e^{-\eps\int_0^{u} p^+(X(s)) \,ds} \left (1-e^{-\eps\int_{u}^{s_i} p^+(X(s)) \,d
s}\right )\, du \right )\\
= \eps^2 \E \left( \int_{s_{i-1}+E_i}^{s_i}  p^+(X(u)) \int_{u}^{s_i} p^+(X(s)) \,ds \, du
 \right)+o(\eps^2)\\
= \eps^2 \E\left(\int_{0\leq u\leq v\leq \wT} \E_\nu\left(p^+(X(0))p^+(X(v-u))\right)\,du\right)  +o(\eps^2),
\end{multline*}
which gives the second expansion.
\end{proof}

The complete expansion is now detailed. 
\begin{prop}\label{prop2}
The  coefficients of $\eps^2$ in  the expansion of $\E(\Delta^\eps)$ are given by
\begin{multline}\label{exp-2}
a_+ = - \frac{\rho}{\mu(1-\rho)} \E\left(\int_{0}^{\wB}(\wB-v)\, \E\left[p^+(X(0))p^+(X(v))\right]\,dv\right) \\
- \frac{1-\rho}{2} \E\left(\int_0^{\wB}  (\wB-u)^2  \E\left[p^+(X(0)) p^+(X(u))\right]\,du\right),
\end{multline}
\begin{multline}\label{exp+2} 
a_- = -\frac{1}{\mu^3(1-\rho)}\E\left(\sum_{1\leq i < j \leq N}  p^-(X(D_i))p^-(X(D_j))\right)\\
-\frac{1}{\mu^2}\E\left(\sum_{i=1}^N \sum_{j=1}^{N'}  p^-(X(D_i))p^-(X(\wT + D_{j}')) \left(\wB_1-D_j' + \frac{\mu}{(\mu-\lambda)^2}\right) \right),
\end{multline}
\begin{multline} \label{exp+-}
a_\pm = \frac{1}{\mu} \E\left(\sum_{i=1}^N \int_{0}^{\wT}  p^-(X(D_i))p^+(X(s)) \, ds \left (\frac{\mu}{(\mu-\lambda)^2}+B-D_i\right) \right)\\
+\frac{1}{\mu}\E\left(\sum_{i=1}^N \int_{0}^{\wB_1} p^-(X(D_i))p^+(X(\wT +s )) \left(\wB_1-s +\frac{\lambda}{(\mu-\lambda)^2} \right)\,ds \right)\\
-\frac{1}{\mu}\E\left(\sum_{i=1}^{H}\sum_{k=1}^{N_i} \int_{0}^{A_i} p^+(X(u))p^-(X(D_i^k )) \left( \frac{\lambda}{(\mu-\lambda)^2} + A_i - u\right) \,du \right)
\end{multline}
where $H$ is geometric distributed with parameter $\lambda/(\mu+\lambda)$, $(N_i,D_1^i,\ldots D^i_{N_i})$
denotes respectively the number of departures and the departures times in a busy period
$B_1^i$, and $A_i=B_1^i+E_0+\sum_{k=i+1}^{H} (E_k+B_1^k)$
where $(E_i)$ are i.i.d is exponentially distributed with parameter $\mu+\lambda
$ and $(B_1^i)$ are i.i.d with the same distribution as $\wT$.
\end{prop}


\end{document}